\documentclass[aps,pra,twocolumn,nofootinbib]{revtex4}
\usepackage{amssymb}
\usepackage{graphicx}

\begin{document}
\title{Degrees of concealment and bindingness in quantum bit commitment protocols}
\author{R. W. Spekkens$^{1}$ and T. Rudolph$^{2}$}
\address{$^{1}$Department of Physics, University of Toronto, 60 St. George Street,\\
Toronto, Ontario, Canada, M5S 1A7\\
$^2$Institut f\"ur Experimentalphysik, Universit\"at Wien, Boltzmanngasse 5,%
\\
1090 Vienna, Austria\\
}
%\date{\today}

\begin{abstract}
Although it is impossible for a bit commitment protocol to be both
arbitrarily concealing and arbitrarily binding, it {\em is}
possible for it to be both {\em partially }concealing and {\em
partially }binding. This means that Bob cannot, prior to the
beginning of the unveiling phase, find out everything about the
bit committed, and Alice cannot, through actions taken after the
end of the commitment phase, unveil whatever bit she desires. We
determine upper bounds on the degrees of concealment and
bindingness that can be achieved simultaneously in {\em any }bit
commitment protocol, although it is unknown whether these can be
saturated. We {\em do}, however, determine the maxima of these
quantities in a restricted class of bit commitment protocols,
namely those wherein all the systems that play a role in the
commitment phase are supplied by Alice. We show that these maxima
can be achieved using a protocol that requires Alice to prepare a
pair of systems in an entangled state, submit one of the pair to
Bob at the commitment phase, and the other at the unveiling phase.
Finally, we determine the form of the trade-off that exists
between the degree of concealment and the degree of bindingness
given various assumptions about the purity and dimensionality of
the states used in the protocol.
\end{abstract}

\maketitle

\section{Introduction}

Bit commitment(BC) is a cryptographic primitive involving two mistrustful
parties, Alice and Bob, wherein one seeks to have Alice submit an encoded
bit of information to Bob in such a way that Bob cannot reliably identify
the bit before Alice decodes it for him, and Alice cannot reliably change
the bit after she has submitted it. In other words, Bob is interested in
binding Alice to some commitment, and Alice is interested in concealing this
commitment from Bob. It is well known \cite{Mayers,LC} that a BC protocol
that is both concealing and binding is impossible \cite{Localitycaveat}.
Nonetheless, it {\em is} possible to devise a BC protocol that is both {\em %
partially} concealing and {\em partially} binding, that is, one wherein if
Alice is honest then the probability that Bob can estimate her commitment
correctly is strictly less than 1, and if Bob is honest then the probability
that Alice can unveil whatever bit she desires is strictly less than 1. This
paper addresses the problem of determining the {\em optimal} degrees of
concealment and bindingness that can be achieved simultaneously in quantum
bit commitment protocols.

We establish an upper bound on the degrees of concealment and bindingness
for all BC protocols. It is unclear at this time whether or not this upper
bound can be saturated. Nonetheless, we {\em are} able to provide a
saturable upper bound for a more restricted class of BC protocols, namely
protocols wherein Alice initially holds all of the systems that play a role
in the commitment phase of the protocol. We also introduce a new kind of BC\
protocol that achieves this maximum. The protocol essentially consists of
Alice preparing two systems in an entangled state, submitting one system to
Bob at the commitment phase, and submitting the other system at the
unveiling phase. We show that in such protocols the maximum achievable
degree of bindingness is related in a simple way to the fidelity between the
reduced density operators for the systems held by Bob at the end of the
commitment phase.

BC appears as a primitive in the protocols of many different cryptographic
tasks between mistrustful parties. As such, the kinds of security that can
be achieved in BC has implications for the kinds of security that can be
achieved in these other tasks. In this paper we consider only the
implications of our results to the task of coin tossing \cite
{LCcointossing,Aharonov}.

\section{Degrees of concealment and bindingness}

A bit commitment protocol involves three phases, which are called the
commitment phase, the holding phase and the unveiling phase. During the
commitment phase, Alice and Bob engage in some number of rounds of
communication, with at least one communication from Alice to Bob. The period
after the end of the commitment phase and prior to the beginning of the
unveiling phase is called the holding phase, and may be of arbitrary
duration. During the unveiling phase, there is again some number of rounds
of communication, with at least one communication from Alice to Bob. At the
end of the unveiling phase, an honest Bob performs a measurement that has
three outcomes, labelled $`0$',`$1$' and `fail', corresponding respectively
to Alice unveiling a 0, Alice unveiling a 1 and Alice being caught cheating.
The protocol specifies the sequence of actions an honest Alice performs in
order to commit to a bit $b$, and guarantees that if she follows the actions
for committing a bit $b$ then an honest Bob's measurement at the end of the
unveiling phase yields the outcome $b$ with certainty.

To discuss the security of BC protocols, it is useful to introduce two
quantities which we shall call Alice's {\em control} and Bob's {\em %
information gain}. These quantities are defined under the assumption that
the other party is honest, and depend on the sequence of actions performed
by the party in question. Alice's control is meant to quantify the extent to
which she can influence (after the commitment phase) the outcome of Bob's
measurement beyond what she could accomplish by following the honest
strategy. Bob's information gain is meant to quantify his ability to
estimate Alice's commitment (prior to the unveiling phase) beyond what he
could accomplish by following the honest strategy.

We now present the specific measures of control and information gain which
we make use of in this paper. We assume for simplicity that Bob has no prior
information on which bit Alice has committed, and that Alice is as likely to
wish to unveil a bit $0$ as she is to wish to unveil a bit $1$. We take our
measure of Bob's information gain for the strategy $S^{B},$ which we denote
by $G\left( S^{B}\right) ,$ to be the difference between his probability of
estimating Alice's commitment correctly when he implements $S^{B}$ and when
he is honest,
\[
G\left( S^{B}\right) =P_{\text{E}}(S^{B})-1/2.
\]
We take our measure of Alice's control for the strategy $S^{A},$ which we
denote by $C\left( S^{A}\right) ,$ to be the difference between her
probability of unveiling whatever bit she desires when she implements $S^{A}$
and when she is honest,
\[
C(S^{A})=P_{\text{U}}(S^{A})-1/2.
\]
It follows that $G(S^{B})$ and $C(S^{A})$ vary between $0$ and $1/2.$

We quantify the degrees of concealment and bindingness in a BC protocol by
Bob's maximum information gain and Alice's maximum control$,$ defined
respectively by
\begin{eqnarray*}
G^{\max } &\equiv &\max_{S^{B}}G\left( S^{B}\right) , \\
C^{\max } &\equiv &\max_{S^{A}}C\left( S^{A}\right) .
\end{eqnarray*}
A protocol is said to be {\em partially concealing }if Bob's maximum
information gain is strictly less than complete information gain, $G^{\max
}<1/2;$ it is said to be {\em perfectly concealing }if his information gain
is zero, $G^{\max }=0;$ finally, it is said to be {\em arbitrarily
concealing }or simply {\em concealing }if his information gain can be made
arbitrarily small by increasing the value of a security parameter $N$, that
is, $G^{\max }\le \varepsilon $, where $\varepsilon \rightarrow 0$ as $%
N\rightarrow \infty $\cite{cheatsensitivity}$.$ Similar definitions hold for
the degrees of security against Alice. A protocol is said to be {\em %
partially binding }if Alice's maximal control is strictly less than complete
control, $C^{\max }<1/2;$ it is said to be {\em perfectly binding }if her
control is zero, $C^{\max }=0;$ finally, it is said to be {\em arbitrarily
binding }or simply {\em binding }if her control can be made arbitrarily
small by increasing the value of a security parameter $N$, that is, $C^{\max
}\le \delta $, where $\delta \rightarrow 0$ as $N\rightarrow \infty .$

If a degree of security (such as concealment or bindingness) can be
guaranteed by assuming only the laws of physics (and the integrity of a
party's laboratory), then it is said to hold {\em unconditionally.} In this
paper, we shall only be concerned with unconditional security. Thus, every
time we assign some degree of security (such as concealment or bindingness)
to a protocol, it is implied that the protocol has this feature
unconditionally.

To understand the degree to which a protocol can be made concealing or
binding we must answer the following questions:

\begin{itemize}
\item  What is Bob's maximal information gain, and what strategy achieves
this maximum? That is, find $G^{\text{max}},$ and find $S_{B}^{\text{max}}$
such that $G(S_{B}^{\text{max}})=G^{\text{max}}.$

\item  What is Alice's maximal control, and what strategy achieves this
maximum? That is, find $C^{\text{max}},$ and find $S_{A}^{\text{max}}$ such
that $C(S_{A}^{\text{max}})=C^{\text{max}}.$
\end{itemize}

In another paper\cite{SpekkensRudolph}, we provide answers to these
questions for BC\ protocols that are generalizations of the BB84 BC protocol%
\cite{BB84}. In this paper, we provide the complete solution for a different
type of BC protocol, which we call a {\em purification }BC protocol.

  The above questions involve an optimization over strategies. We
will also be interested in optimizing over protocols. Specifically, we wish
to answer the following question:

\begin{itemize}
\item  For a given class of protocols, what is the {\em minimum }Alice's
maximum control can be made for a given value of Bob's maximum information
gain, and which protocol in the class achieves this minimum? In other words,
denoting protocols by ${\cal P}$, the given class of protocols by ${\cal K},$
and the subset of this class associated with $G^{\max }$ by ${\cal K}%
(G^{\max }),$ find $\min_{{\cal P\in K}(G^{\max })}C^{\max }\left( {\cal P}%
\right) $ and find ${\cal P}^{\text{opt}}$ such that $C^{\max }\left( {\cal P%
}^{\text{opt}}\right) =\min_{{\cal P\in K}(G^{\max })}C^{\max }\left( {\cal P%
}\right) $.
\end{itemize}

If this question can be answered for every value of $G^{\max },$ then one
obtains a curve in the $G^{\max }$-$C^{\max }$ plane. Moreover, if this
curve is monotonically decreasing then it is identical to what would have
been obtained by minimizing Bob's maximum information gain for a given value
of Alice's maximum control. In this case, we call the curve the {\em optimal
trade-off relation} between $C^{\max }$ and $G^{\max }$. Specifying this
relation for a given class of protocols is a convenient way of expressing
the maximum degrees of concealment and bindingness that can be achieved with
such protocols.

In this paper, we determine a lower bound on the optimal trade-off relation
between $C^{\max }$ and $G^{\max }$ for all BC\ protocols. Unfortunately, we
have not determined whether this lower bound is saturable or not. However,
we do find the optimal trade-off relation for a restricted class of BC
protocols, which we call{\em \ Alice-supplied} BC\ protocols. The
generalized BB84 BC protocols and the purification BC protocols mentioned
above both fall into this class. In fact, we show that the purification BC
protocols are optimal within this class. These protocols will be defined
precisely in the next section.

\section{BC protocols}

  In order to perform optimizations over all quantum BC protocols,
it is necessary to have a completely general model of such protocols. We
make use of the following model for cryptographic protocols implemented
between two mistrustful parties\cite{Mayers}. The Hilbert space required to
describe the protocol is the tensor product of the Hilbert spaces for all
the systems that play a role in the protocol. Every action taken by a party
in their laboratory corresponds to that party performing a unitary operation
on the systems in their possession. Every communication corresponds to a
party sending some subset of the systems in their possession to the other
party (it follows that the mere transmission of information from one party
to the other does not change the quantum state of the total system, but does
change the Hilbert space upon which the parties can implement their unitary
operations). It is assumed that the total system is initially in a pure state%
$.$

It has been previously argued\cite{Mayers} that this model is completely
general. It incorporates the possibility of random choices and measurements
during the protocol, since these can always be kept at the quantum level
until the end without any loss of generality. A random choice is performed
at the quantum level by implementing a unitary transformation that is
conditioned upon the state of an ancilla prepared initially in a
superposition of states. Measurements are performed at the quantum level by
unitarily coupling the system to be measured to an ancilla that is prepared
in some fixed initial pure state.

In the case of BC, the most general protocol may involve many rounds of
communication during the commitment phase. Denoting the number of rounds by $%
n,$ denoting Alice's honest sequence of operations for committing a bit $b$
by $\{W_{b,1},...,W_{b,n}\},$ and denoting Bob's honest sequence of
operations by $\{W_{1}^{\prime },...,W_{n}^{\prime }\},$ the total unitary
operation they jointly implement is
\[
W_{b}\equiv W_{n}^{\prime }W_{b,n}\cdots W_{2}^{\prime }W_{b,2}W_{1}^{\prime
}W_{b,1}.
\]
The transmissions that occur in each round will determine the Hilbert space
over which $W_{b,i}$ and $W_{i}^{\prime }$ act non-trivially. Thus, despite
the fact that we have assumed that Alice implements the first unitary
operation, this operation could be trivial and it remains arbitrary which
party is first to submit a sytem to the other party. If the initial state of
all systems is denoted by $\left| \psi _{\text{init}}\right\rangle , $ then
the state at the holding phase if both parties are honest is
\[
\left| \psi _{b}\right\rangle \equiv W_{b}\left| \psi _{\text{init}%
}\right\rangle .
\]
It follows that the reduced density operator for Bob's system at the holding
phase, assuming both parties are honest, is
\[
\rho _{b}=Tr\left( \left| \psi _{b}\right\rangle \left\langle \psi
_{b}\right| \right) ,
\]
where the trace is over all the systems that end up in Alice's possession at
the holding phase.

During the unveiling phase, a similar process occurs. Denoting the number of
rounds by $m,$ denoting Alice's honest sequence of operations given that she
committed to bit $b$ by $\{V_{b,1},...,V_{b,n}\},$ and denoting Bob's honest
sequence of operations by $\{V_{1}^{\prime },...,V_{n}^{\prime }\},$ the
total unitary operation they jointly implement is
\[
V_{b}\equiv V_{n}^{\prime }V_{b,n}\cdots V_{2}^{\prime }V_{b,2}V_{1}^{\prime
}V_{b,1}.
\]
Thus, if both parties are honest, the state of the total system at the end
of the unveiling phase is
\begin{equation}
\left| \psi _{b}^{\text{unv}}\right\rangle \equiv V_{b}\left| \psi
_{b}\right\rangle .  \label{psibunv}
\end{equation}
The protocol ends with Bob performing a three-outcome projective measurement
$\left\{ \Pi _{0},\Pi _{1},\Pi _{\text{fail}}\right\} $ on the systems in
his possession. If both parties are honest, then whenever Alice commits to a
bit $b,$ the measurement must have outcome $b$ with probability $1.$ This
implies that $\left| \psi _{0}^{\text{unv}}\right\rangle $ and $\left| \psi
_{1}^{\text{unv}}\right\rangle $ must be orthogonal,
\[
\left\langle \psi _{0}^{\text{unv}}|\psi _{1}^{\text{unv}}\right\rangle =0,
\]
and that $\left| \psi _{b}^{\text{unv}}\right\rangle $ must be an eigenstate
of $\Pi _{b}$ with eigenvalue $1,$%
\begin{equation}
\Pi _{b}\left| \psi _{b}^{\text{unv}}\right\rangle =\left| \psi _{b}^{\text{%
unv}}\right\rangle .  \label{conditionofcorrectness}
\end{equation}

As mentioned earlier, we will be interested in a restricted class of BC
protocols, which we call {\em Alice-supplied }BC protocols{\em . }These
protocols impose no restrictions on the details of the unveiling phase and
may involve an arbitrary number of rounds of communication between Alice and
Bob during the commitment phase. However, it is required that {\em all} of
the systems that Bob makes use of during the commitment phase are supplied
by Alice. The class of Alice-supplied BC protocols includes the generalized
BB84 BC protocols, defined in Ref. \cite{SpekkensRudolph}, as well as the
purification BC protocols defined below. An example of a protocol that falls
{\em outside }this class is one wherein at the beginning of the commitment
phase Bob submits to Alice a system that is entangled with one he keeps in
his possession, and Alice encodes her commitment in the unitary
transformation she performs upon this system before resubmitting it to Bob.
Another example of such a protocol is one wherein during the commitment
phase Bob uses ancillas that Alice did not supply in order to make a random
choice or perform a measurement.

We now provide a precise definition of a purification BC protocol.

{\bf A purification BC protocol. }Such a protocol makes use of just two
systems, which we shall call the token system and the proof system (since
one is the token of Alice's commitment and the other is the proof of her
commitment). These are associated with Hilbert spaces ${\cal H}_{p}$ and $%
{\cal H}_{t}.$ A purification BC protocol also specifies two orthogonal
states $\left| \chi _{0}\right\rangle $ and $\left| \chi _{1}\right\rangle $
defined on ${\cal H}_{p}\otimes {\cal H}_{t}$. The honest actions are as
follows.

\begin{enumerate}
\item  At the commitment phase, Alice prepares the two systems in the state $%
\left| \chi _{b}\right\rangle $ in order to commit to bit $b,$ and sends the
token system to Bob.

\item  At the unveiling phase, Alice sends the proof system to Bob, and Bob
performs a measurement of the projector-valued measure $\left\{ \Pi _{0},\Pi
_{1},\Pi _{\text{fail}}\right\} ,$ where $\Pi _{b}=\left| \chi
_{b}\right\rangle \left\langle \chi _{b}\right| .$
\end{enumerate}

So we see there is only a single communication from Alice to Bob during both
the commitment and the unveiling phases. In the notation of the general
model presented above, $W_{b}$ transforms $\left| \psi _{\text{init}%
}\right\rangle $ to $\left| \psi _{b}\right\rangle =\left| \chi
_{b}\right\rangle ,$ and $V_{b}=I$ so that $\left| \psi _{b}^{\text{unv}%
}\right\rangle =\left| \psi _{b}\right\rangle =\left| \chi _{b}\right\rangle
.$

We call this a purification BC protocol, since at the unveiling phase an
honest Alice is required to provide Bob with a purification of the state
that he received from her during the commitment phase.

\section{\protect  Measures of distinguishability for density
operators}

Two measures of the distinguishability of density operators will be
important in the present work: the trace distance and the fidelity, defined
respectively by\cite{NielsenChuang}
\[
D\left( \rho ,\sigma \right) =\frac{1}{2}Tr\left| \rho -\sigma \right| ,
\]
and
\[
F\left( \rho ,\sigma \right) =Tr\left| \sqrt{\rho }\sqrt{\sigma }\right| ,
\]
where $\left| A\right| =\sqrt{A^{\dag }A}$.

  We will find the following relations between these two measures
to be very useful. For any two density operators, the fidelity and the trace
distance satisfy\cite{FuchsGraaf}
\begin{equation}
1-F\left( \rho ,\sigma \right) \le D(\rho ,\sigma ),  \label{FDrelations}
\end{equation}
and
\begin{equation}
D(\rho ,\sigma )\le \sqrt{1-F(\rho ,\sigma )^{2}}.  \label{FDrelation2}
\end{equation}
The second inequality is saturated for any pair of pure states, that is,
\begin{equation}
D(\left| \psi \right\rangle ,\left| \chi \right\rangle )=\sqrt{1-F(\left|
\psi \right\rangle ,\left| \chi \right\rangle )^{2}},
\label{FDrelation4pure}
\end{equation}
for all $\left| \psi \right\rangle $ and $\left| \chi \right\rangle .$ A
stronger lower bound for the trace distance between $\rho $ and $\sigma $
exists if one of the density operators is pure. Specifically,
\begin{equation}
1-F\left( \rho ,\left| \psi \right\rangle \right)^{2} \le D(\rho ,\left|
\psi \right\rangle ).  \label{FDrelation4onepure}
\end{equation}
This stronger lower bound also applies to the mixed states of qubits. More
precisely, we have the following result.

\begin{description}
\item[Lemma 1]  For pairs of density operators $\rho ,$ $\sigma $ whose
supports lie in a single 2-dimensional Hilbert space,
\[
1-F\left( \rho ,\sigma \right) ^{2}\le D(\rho ,\sigma ).
\]
\end{description}

The proof of this is presented in the appendix. All of the above
inequalities can be saturated. Explicit examples will be presented in
section VI.

Finally, we present some properties of the fidelity that will be useful for
the present investigation. Uhlmann's theorem\cite{Jozsa} states that the
fidelity between two density operators is equal to the overlap of two
maximally parallel purifications of these density operators. Thus, if $\rho $
and $\sigma $ are density operators on a Hilbert space ${\cal H},$ $\left|
\psi \right\rangle $ and $\left| \chi \right\rangle $ are arbitrary
purifications of $\rho $ and $\sigma $ on ${\cal H}^{\prime }\otimes {\cal H}%
,$ and $U$ is a unitary transformation on ${\cal H}^{\prime },$ then
\begin{equation}
F\left( \rho ,\sigma \right) =\max_{U}\left| \left\langle \psi \right|
U\otimes I\left| \chi \right\rangle \right| .  \label{Uhlmann}
\end{equation}

Another critical property is given by the following lemma.

\begin{description}
\item[Lemma 2]  The fidelity satisfies
\[
\max_{\rho }\left( F\left( \rho ,\sigma \right) ^{2}+F\left( \rho ,\omega
\right) ^{2}\right) =1+F\left( \sigma ,\omega \right) .
\]
\end{description}

The proof of this can be found in the derivation of Eq.(\ref{Pumax4}) from
Eq.(\ref{Pumax3}) in section VI and by making use of Uhlmann's theorem.

\section{Optimizing over all BC protocols}

In this section, we demonstrate an upper bound on the simultaneous degrees
of concealment and bindingness (hence a {\em lower} bound on $G^{\max }$ and
$C^{\max }$) for {\em any }BC protocol. It should be noted that the main
ideas that go into the proof of this result are present in the work of Mayers%
\cite{Mayers} and Lo and Chau\cite{LCcointossing}.

\begin{description}
\item[Theorem 1]  In any BC protocol,
\begin{eqnarray*}
\text{i) }G^{\max } &\ge &\frac{1}{2}D\left( \rho _{0},\rho _{1}\right) , \\
\text{ii) }C^{\max } &\ge &\frac{1}{2}F\left( \rho _{0},\rho _{1}\right)
^{2}.
\end{eqnarray*}
\end{description}

{\bf Proof}. We begin by proving (i). To analyze security against Bob, we
assume that Alice is honest. Suppose that Bob uses a strategy wherein he
acts honestly throughout the commitment phase. In this case, the state of
the total system at the end of this phase will be $\left| \psi
_{0}\right\rangle $ or $\left| \psi _{1}\right\rangle ,$ depending on
Alice's commitment. The reduced density operators for Bob's system will be $%
\rho _{0}$ or $\rho _{1}.$ Now suppose that during the holding phase Bob
does the measurement which optimally discriminates between $\rho _{0}$ and $%
\rho _{1}$. It is a well-known result of state estimation theory\cite
{Helstrom,Fuchs} that his information gain in this case will be
\[
G=\frac{1}{2}D\left( \rho _{0},\rho _{1}\right) .
\]
Bob's maximum information gain may be greater than this value, since it may
be beneficial for him to also cheat during the commitment phase (for
instance, if the reduced density operators on Bob's systems are more easily
discriminated at some point during the commitment phase than they are at the
holding phase). Bob's maximum information gain cannot, however, be less than
this bound. This establishes (i).

We now prove (ii). To analyze security against Alice, we can assume that Bob
is honest. Suppose that Alice uses the following strategy. During the
commitment phase, she follows the honest protocol for committing a bit $0$,
so that the total system is in the state $\left| \psi _{0}\right\rangle $ at
the holding phase. Thereafter, if Alice wishes to unveil a bit $0,$ she acts
honestly for the rest of the protocol, while if she wishes to unveil a bit $%
1,$ then she applies a unitary transformation $U^{\max }$ to the systems in
her possession just prior to the unveiling phase, and thereafter acts
honestly. $U^{\max }$ is chosen such that
\begin{equation}
\left\langle \psi _{1}\right| U^{\max }\otimes I\left| \psi
_{0}\right\rangle =\max_{U}\left\langle \psi _{1}\right| U\otimes I\left|
\psi _{0}\right\rangle .  \label{defineU}
\end{equation}
The probability that Alice succeeds at unveiling a bit $0$ when she attempts
to do so is unity, $P_{\text{U}0}=1,$ since she has simply followed the
honest protocol for committing a $0.$ The probability that Alice succeeds at
unveiling a bit $1$ when she attempts to do so is
\[
P_{\text{U}1}=Tr\left( \Pi _{1}V_{1}\left( U^{\max }\otimes I\right) \left|
\psi _{0}\right\rangle \left\langle \psi _{0}\right| \left( U^{\max \dag
}\otimes I\right) V_{1}^{\dag }\right) .
\]
Now since the state $\left| \psi _{1}^{\text{unv}}\right\rangle =V_{1}\left|
\psi _{1}\right\rangle $ is an eigenstate of $\Pi _{1}$ with eigenvalue 1
(see Eq.(\ref{conditionofcorrectness})), one can write
\[
\Pi _{1}=\left| \psi _{1}^{\text{unv}}\right\rangle \left\langle \psi _{1}^{%
\text{unv}}\right| +\Gamma _{1},
\]
for some non-negative operator $\Gamma _{1}$, orthogonal to $\left| \psi
_{1}^{\text{unv}}\right\rangle \left\langle \psi _{1}^{\text{unv}}\right| .$
It follows that
\begin{eqnarray*}
P_{\text{U}1} &\ge &\left| \left\langle \psi _{1}^{\text{unv}}\right|
V_{1}\left( U^{\max }\otimes I\right) \left| \psi _{0}\right\rangle \right|
^{2} \\
&=&\left| \left\langle \psi _{1}\right| U^{\max }\otimes I\left| \psi
_{0}\right\rangle \right| ^{2}.
\end{eqnarray*}
Since we are assuming that Alice is equally likely to wish to unveil a $0$
as a $1,$ her probability of unveiling the bit of her choosing satisfies
\begin{eqnarray*}
P_{\text{U}} &=&\frac{1}{2}P_{\text{U}0}+\frac{1}{2}P_{\text{U}1} \\
&\ge &\frac{1}{2}+\frac{1}{2}\left| \left\langle \psi _{1}\right| U^{\max
}\otimes I\left| \psi _{0}\right\rangle \right| ^{2}.
\end{eqnarray*}
Recalling the definition of $U^{\max }$ (Eq.(\ref{defineU})), and making use
of Uhlmann's theorem (Eq.(\ref{Uhlmann})), we conclude that Alice's control
for this particular strategy satisfies
\[
C\ge \frac{1}{2}F\left( \rho _{0},\rho _{1}\right) ^{2}.
\]
Alice's maximum control may be greater than this bound, since she may be
able to cheat during the commitment and unveiling phases as well, but it
cannot be less. This establishes (ii). $\Box $

\begin{figure}
\includegraphics[width=80mm]{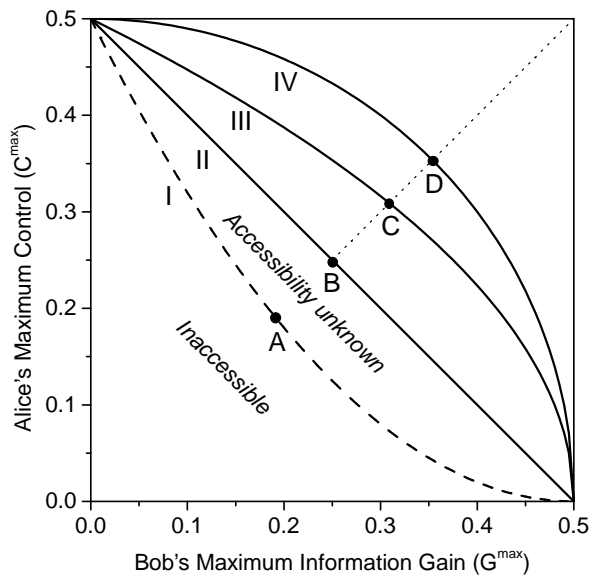}
\caption{Curve I is a lower bound for the trade-off relation
between $C^{\max }$ and $G^{\max }$ for {\em any} BC protocol. The
other curves are the optimal trade-off relations for
Alice-supplied BC\ under different restrictions on $\rho _{0}$ and
$\rho _{1}$: (II) no restrictions; (III) both qubit states or not
both mixed states; and (IV) both pure states. A, B, C and D
correspond to the points along these curves where the protocol is
fair, i.e. $C^{\max }=G^{\max }.$}
\end{figure}

It is common in quantum information theory to question the degree to which
the sharing of prior entanglement enhances one's ability to perform
information processing tasks. With this in mind it is perhaps interesting to
note that the proof of Theorem 1 makes no restriction on $|\psi _{\text{init}%
}\rangle $. Thus theorem 1 applies even if Alice and Bob share entangled
states that they both trust prior to the initialization of the BC protocol.

\begin{description}
\item[Corollary 1]  In any BC protocol, the optimal trade-off between $%
G^{\max }$ and $C^{\max }$ is a curve satisfying
\[
2G^{\max }+\sqrt{2C^{\max }}\ge 1.
\]
(the lower bound corresponds to curve I in Fig. 1).
\end{description}

{\bf Proof: }This follows from Theorem 1 and Eq. (\ref{FDrelations}). $\Box
\medskip $

It is well-known\cite{Mayers,LC} that it is impossible to have a BC\
protocol that is both arbitrarily concealing and arbitrarily binding, that
is, one for which $G^{\max }\le \varepsilon $ and $C^{\max }\le \delta $ for
arbitrarily small $\varepsilon $ and $\delta .$ This clearly follows from
Corollary 1. However, Corollary 1 says {\em more} than this, since it also
sets a lower bound on the extent to which any BC\ protocol can be partially
concealing and partially binding. Thus, in addition to being able to rule
out the possibility of a BC\ protocol with $G^{\max }$ and $C^{\max }$
arbitrarily close to the origin in Fig. 1, one can rule out the possibility
of a BC protocol anywhere below curve I of Fig. 1. The best one can hope for
is a BC\ protocol with $2G^{\max }+\sqrt{2C^{\max }}=1$ (curve I of Fig. 1).
In particular, the best fair BC\ protocol one can hope for has $C^{\max
}=G^{\max }=\frac{3-\sqrt{5}}{4}\approx .19098$ (point A in Fig. 1).

In this paper, we do not settle the question of whether there exists a
protocol for which Alice's maximal control and Bob's maximal information
gain achieve the lower bounds of Theorem 1 simultaneously. Such a protocol
would have to be such that Bob could not get any more information by
cheating during the commitment phase than he can by cheating during the
holding phase, and such that Alice could not get any more control by
cheating during the commitment phase or the unveiling phase than she can by
cheating during the holding phase. It seems to us that such a protocol is
unlikely to exist.

\section{Optimizing over Alice-supplied BC protocols}

\subsection{Optimal degrees of concealment and bindingness}

The main results of this paper are:

\begin{description}
\item[Theorem 2]  In Alice-supplied BC protocols,
\begin{eqnarray*}
\text{i) }G^{\max } &\ge &\frac{1}{2}D\left( \rho _{0},\rho _{1}\right) , \\
\text{ii) }C^{\max } &\ge &\frac{1}{2}F\left( \rho _{0},\rho _{1}\right) ,
\end{eqnarray*}
and

\item[Theorem 3]  Purification BC protocols saturate the bounds in Theorem
2.\bigskip
\end{description}

{\bf Proof of Theorem 2. }Inequality (i) follows trivially from theorem 1,
since if $G^{\max }\ge \frac{1}{2}D\left( \rho _{0},\rho _{1}\right) $ for
{\em all} BC protocols then clearly $G^{\max }\ge \frac{1}{2}D\left( \rho
_{0},\rho _{1}\right) $ for any Alice-supplied BC protocol.

  Inequality (ii), on the other hand, is stronger than theorem 1.
To prove it, we must consider Alice's most general cheating strategy.
Without loss of generality, we can assume that she keeps all of her cheating
actions at the quantum level. During the commitment phase, Alice can cheat
by implementing a sequence of unitary operations $\{\tilde{W}_{1},\dots ,%
\tilde{W}_{n}\}$ different from the honest sequence$.$ She can cheat at the
end of the holding phase by implementing a unitary transformation $%
U_{b}\otimes I$ that depends on the bit $b$ she would like to unveil.
Finally, she can cheat during the unveiling phase by implementing a sequence
of unitary operations $\{\tilde{V}_{b,1},\dots ,\tilde{V}_{b,n}\}$ that
depends on the bit $b$ she would like to unveil and that is different from
the honest sequence$.$ The maximum probability of Alice unveiling the bit of
her choosing is therefore given by
\begin{eqnarray*}
P_{\text{U}}^{\max } &=&\frac{1}{2}\max_{\{\tilde{W}_{1},\dots ,\tilde{W}%
_{n}\}}\sum_{b\in \{0,1\}}\max_{\{\tilde{V}_{b,1},\dots ,\tilde{V}%
_{b,n}\}}\max_{U_{b}} \\
&&Tr\{\Pi _{b}\tilde{V}_{b}\left( U_{b}\otimes I\right) \tilde{W}\left| \psi
_{\text{init}}\right\rangle \left\langle \psi _{\text{init}}\right| \\
&&\times \tilde{W}^{\dag }\left( U_{b}^{\dag }\otimes I\right) \tilde{V}%
_{b}^{\dag }\},
\end{eqnarray*}
where
\begin{eqnarray*}
\tilde{W} &\equiv &W_{n}^{\prime }\tilde{W}_{n}\cdots W_{2}^{\prime }\tilde{W%
}_{2}W_{1}^{\prime }\tilde{W}_{1},\text{ and} \\
\tilde{V}_{b} &\equiv &V_{n}^{\prime }\tilde{V}_{b,n}\cdots V_{2}^{\prime }%
\tilde{V}_{b,2}V_{1}^{\prime }\tilde{V}_{b,1}.
\end{eqnarray*}
$\tilde{W}$ and $\tilde{V}_{b}$ are the total unitary operations that Alice
and Bob jointly implement given that Bob is honest and Alice cheats.

We begin by optimizing over Alice's cheating strategy during the
commitment phase. It turns out that the assumption of an
Alice-supplied protocol allows us to replace the maximization over
$\{\tilde{W}_{1},\dots ,\tilde{W}_{n}\}$ by a maximization over
{\em all} unitary operations on the total system. This means that
Alice has as much cheating power in an arbitrary Alice-supplied
protocol as she does in a protocol where Bob does not play any
role in the commitment phase.   The reason is that Alice can bring
about any unitary operation $W$ by implementing the sequence of
operations
\begin{eqnarray*}
\tilde{W}_{1} &=&\left( W_{n}^{\prime }\cdots W_{1}^{\prime }\right) ^{-1}W
\\
\tilde{W}_{i} &=&I\text{ for }i\ne 1.
\end{eqnarray*}
This result only applies for Alice-supplied BC protocols, since Alice must
initially have access to all the systems that will appear in the commitment
phase in order to implement $\tilde{W}_{1}.$ We can conclude that
\begin{eqnarray*}
P_{\text{U}}^{\max } &=&\frac{1}{2}\max_{W}\sum_{b\in \{0,1\}}\max_{\{\tilde{%
V}_{b,1},\dots ,\tilde{V}_{b,n}\}}\max_{U_{b}} \\
&&Tr(\Pi _{b}\tilde{V}_{b}\left( U_{b}\otimes I\right) W\left| \psi _{\text{%
init}}\right\rangle \left\langle \psi _{\text{init}}\right| \\
&&\times W^{\dag }\left( U_{b}^{\dag }\otimes I\right) \tilde{V}_{b}^{\dag
}).
\end{eqnarray*}

We now consider the unveiling measurement. Eq.(\ref{conditionofcorrectness})
implies that the honest state at the end of the unveiling phase, $\left|
\psi _{b}^{\text{unv}}\right\rangle $ must be an eigenstate of $\Pi _{b}.$
Thus,
\[
\Pi _{b}=\left| \psi _{b}^{\text{unv}}\right\rangle \left\langle \psi _{b}^{%
\text{unv}}\right| +\Gamma _{b},
\]
for some non-negative operator $\Gamma _{b}$. It follows that
\begin{eqnarray*}
P_{\text{U}}^{\max } &\ge &\frac{1}{2}\max_{W}\sum_{b\in \{0,1\}}\max_{\{%
\tilde{V}_{b,1},\dots ,\tilde{V}_{b,n}\}}\max_{U_{b}} \\
&&\left| \left\langle \psi _{b}^{\text{unv}}|\tilde{V}_{b}\left(
U_{b}\otimes I\right) W|\psi _{\text{init}}\right\rangle \right| ^{2}.
\end{eqnarray*}

Clearly the maximum over $\{\tilde{V}_{b,1},\dots ,\tilde{V}_{b,n}\}$ must
be greater than or equal to the value for $\{V_{b,1},\dots ,V_{b,n}\}$, the
honest sequence of operations for unveiling bit $b.$ Thus,
\[
P_{\text{U}}^{\max }\ge \frac{1}{2}\max_{W}\sum_{b\in
\{0,1\}}\max_{U_{b}}\left| \left\langle \psi _{b}^{\text{unv}}|V_{b}\left(
U_{b}\otimes I\right) W|\psi _{\text{init}}\right\rangle \right| ^{2}.
\]
Since $W$ varies over all unitary operators, we can write $\left| \psi
\right\rangle =W\left| \psi _{\text{init}}\right\rangle $ and vary over all $%
\left| \psi \right\rangle .$ Making use of the fact that $\left| \psi _{b}^{%
\text{unv}}\right\rangle =V_{b}\left| \psi _{b}\right\rangle $
(Eq.(\ref {psibunv})),   we have
\begin{equation}
P_{\text{U}}^{\max }\ge \frac{1}{2}\max_{\left| \psi \right\rangle
}\sum_{b\in \{0,1\}}\max_{U_{b}}\left| \left\langle \psi _{b}|\left(
U_{b}\otimes I\right) |\psi \right\rangle \right| ^{2}.  \label{Pumax3}
\end{equation}
We perform the maximization over $\left| \psi \right\rangle $ for a given $%
U_{0}$ and $U_{1}.$ By a variational approach, it is easy to show that the
optimal $\left| \psi \right\rangle $ has the form (up to an arbitrary
overall phase)
\begin{equation}
\left| \psi ^{\text{max}}\right\rangle =\frac{\left| \tilde{\psi}%
_{0}\right\rangle +e^{-i\arg \left( \left\langle \tilde{\psi}_{0}|\tilde{\psi%
}_{1}\right\rangle \right) }\left| \tilde{\psi}_{1}\right\rangle }{\sqrt{2}%
\sqrt{1+\left| \left\langle \tilde{\psi}_{0}|\tilde{\psi}_{1}\right\rangle
\right| }},  \label{psimax}
\end{equation}
where
\begin{eqnarray*}
\left| \tilde{\psi}_{0}\right\rangle &=&\left( U_{0}\otimes I\right) \left|
\psi _{0}\right\rangle \\
\left| \tilde{\psi}_{1}\right\rangle &=&\left( U_{1}\otimes I\right) \left|
\psi _{1}\right\rangle .
\end{eqnarray*}
It follows that
\begin{equation}
P_{\text{U}}^{\max }\ge \frac{1}{2}\left( 1+\max_{U_{0},U_{1}}\left|
\left\langle \psi _{0}\right| U_{0}U_{1}\otimes I\left| \psi
_{1}\right\rangle \right| \right) .  \label{Pumax4}
\end{equation}
Inequality (ii) now follows trivially from Uhlmann's theorem and the
definition of Alice's control.$\Box $ \bigskip

{\bf Proof of Theorem 3. }  Recall the definition of a
purification BC protocol, provided in section III. If Alice is
honest she prepares the proof-token composite in either $\left|
\chi _{0}\right\rangle $ or $\left| \chi _{1}\right\rangle $ and
submits the token system to Bob. In this case, the reduced density
operators $\rho _{0}$ and $\rho _{1}$ that describe the token
system are simply the trace over the proof system of $\left| \chi
_{0}\right\rangle $ and $\left| \chi _{1}\right\rangle $, that is,
\[
\rho _{b}=Tr_{p}\left( \left| \chi _{b}\right\rangle \left\langle \chi
_{b}\right| \right) .
\]
The only cheating strategy available to Bob is to try to estimate the state
of the token system, that is, to discriminate $\rho _{0}$ and $\rho _{1}.$
It follows from state estimation theory that his maximum information gain is
$G^{\max }=\frac{1}{2}D\left( \rho _{0},\rho _{1}\right) $ and is achieved
by performing a Helstrom measurement\cite{Helstrom,Fuchs}.

Alice can cheat in {\em two} ways in a purification BC protocol. She can
cheat during the commitment phase by preparing the total system in a state $%
\left| \psi \right\rangle $ that is different from $\left| \chi
_{0}\right\rangle $ or $\left| \chi _{1}\right\rangle ,$ and she can cheat
just prior to the unveiling phase by implementing a unitary operation $U_{b}$
on the proof system. The identity of $U_{b}$ can of course depend on which
bit $b$ she wishes to unveil.

Recalling that $\Pi _{b}=\left| \chi _{b}\right\rangle \left\langle \chi
_{b}\right| ,$ Alice's maximum probability of unveiling whatever bit she
desires is

\[
P_{\text{U}}^{\max }=\max_{\left| \psi \right\rangle }\sum_{b\in \left\{
0,1\right\} }\frac{1}{2}\max_{U_{b}}\left| \left\langle \chi
_{b}|U_{b}\otimes I|\psi \right\rangle \right| ^{2}.
\]
Defining $\rho \equiv Tr_{p}\left( \left| \psi \right\rangle \left\langle
\psi \right| \right) $ and making use of Uhlmann's theorem, we obtain
\[
P_{\text{U}}^{\max }=\frac{1}{2}\max_{\rho }\left( F\left( \rho ,\rho
_{0}\right) ^{2}+F\left( \rho ,\rho _{0}\right) ^{2}\right) .
\]
It now follows trivially from Lemma 2 and the definition of the control that
$C^{\max }=\frac{1}{2}F\left( \rho _{0},\rho _{1}\right) $. Alice achieves
this control by implementing any unitary operations $U_{0}$ and $U_{1}$ that
satisfy $U_{0}U_{1}=U^{\max }$ where $U^{\max }$ is defined in Eq.(\ref
{defineU}), and by initially preparing the state $\left| \psi ^{\max
}\right\rangle $ of Eq.(\ref{psimax}) with $\left| \psi _{b}\right\rangle
=\left| \chi _{b}\right\rangle .$ $\Box $

\subsection{Optimal trade-off relations}

  Given theorem 3, it is straightforward to determine the optimal
trade-off relations between $G^{\max }$ and $C^{\max }$ for various
restrictions on the states of Bob's system at the holding phase.

\begin{description}
\item[Corollary 2]  In Alice-supplied BC\ protocols where $\rho _{0}$ and $%
\rho _{1}$ are arbitrary, the optimal trade-off is
\[
G^{\max }+C^{\max }=\frac{1}{2}
\]
(This corresponds to curve II in Fig. 1).
\end{description}

{\bf Proof. }This follows from theorem 3 and Eq.(\ref{FDrelations}). $\Box $

\begin{description}
\item[Corollary 3]  In Alice-supplied BC protocols where $\rho _{0}$ and $%
\rho _{1}$ either (1) have supports that lie in a single 2 dimensional
Hilbert space, or (2) are not both mixed, the optimal trade-off is
\[
G^{\max }+2(C^{\max })^{2}=\frac{1}{2}.
\]
(This corresponds to curve III in Fig. 1).
\end{description}

{\bf Proof. }This follows from theorem 3, Eq.(\ref{FDrelation4onepure}) and
Lemma 1. $\Box $

\begin{description}
\item[Corollary 4]  In Alice-supplied BC protocols where $\rho _{0}$ and $%
\rho _{1}$ are both pure states,{\em \ }the optimal trade-off is
\[
\left( G^{\max }\right) ^{2}+\left( C^{\max }\right) ^{2}=\frac{1}{4}.
\]
(This corresponds to curve IV in Fig. 1).
\end{description}

{\bf Proof.} This follows from theorem 3 and Eq.(\ref{FDrelation4pure}). $%
\Box \medskip $

  We now provide simple examples of protocols that achieve the
optimal trade-offs of Corollaries 2-4.

To achieve the optimal trade-off of Corollary 2, it suffices to consider a
purification BC protocol where $\rho _{0}$ and $\rho _{1}$ saturate the
inequality of Eq.(\ref{FDrelations}). The simplest example makes use of
commuting density operators in a 3 dimensional Hilbert space. Specifically,
\[
\rho _{0}=\left(
\begin{array}{lll}
\lambda & 0 & 0 \\
0 & 1-\lambda & 0 \\
0 & 0 & 0
\end{array}
\right) \text{ and }\rho _{1}=\left(
\begin{array}{lll}
0 & 0 & 0 \\
0 & 1-\lambda & 0 \\
0 & 0 & \lambda
\end{array}
\right) .
\]
It is straightforward to show that $D\left( \rho _{0},\rho _{1}\right)
=\lambda $ and $F\left( \rho _{0},\rho _{1}\right) =1-\lambda ,$ which
implies that $D\left( \rho _{0},\rho _{1}\right) +F\left( \rho _{0},\rho
_{1}\right) =1$. It is worth emphasizing that a 3 dimensional Hilbert space
is the smallest space in which this bound can be saturated, since states in
a 2 dimensional Hilbert space must satisfy lemma 1.

  We now provide a specific example of a family of protocols that
achieve the optimal trade-off of Corollary 3. We consider purification BC\
protocols wherein
\[
\rho _{0}=\left(
\begin{array}{ll}
1 & 0 \\
0 & 0
\end{array}
\right) \text{ and }\rho _{1}=\left(
\begin{array}{ll}
\lambda  & 0 \\
0 & 1-\lambda
\end{array}
\right) .
\]
Note that this example qualifies both as an example where $\rho _{0}$ and $%
\rho _{1}$ have supports that lie in the same 2 dimensional Hilbert space,
and as an example where one of $\rho _{0}$ and $\rho _{1}$ is pure. It is
easy to see that $D\left( \rho _{0},\rho _{1}\right) =1-\lambda $ and $%
F\left( \rho _{0},\rho _{1}\right) =\sqrt{\lambda }.$ Thus, we have
saturated the lower bounds in Eq.(\ref{FDrelation4onepure}) and lemma 1, and
consequently, this family of protocols is optimal for the specified
restrictions on $\rho _{0}$ and $\rho _{1}$.

It is trivial to find BC\ protocols that achieve the optimal trade-off of
Corollary 4. Any purification BC\ protocol where $\rho _{0}$ and $\rho _{1}$
are pure states will do. Specifically, if
\[
\rho _{0}=\left| 0\right\rangle \left\langle 0\right| \text{ and }\rho
_{1}=\left| \phi \right\rangle \left\langle \phi \right| ,
\]
where $\left| \phi \right\rangle =\cos \phi \left| 0\right\rangle +\sin \phi
\left| 1\right\rangle ,$ then one achieves every point on the curve $\left(
C^{\max }\right) ^{2}+\left( G^{\max }\right) ^{2}=\frac{1}{4}$ by varying
over $\phi $ in the range $0$ to $\pi /2.$

  If we define a `fair' BC protocol to be one where $C^{\max
}=G^{\max },$ then by substituting this identity into the trade-off
relations presented above, we obtain the following results. The best fair BC
protocol from among the class of Alice-supplied BC\ protocols has $C^{\max
}=G^{\max }=0.25$ (point A on Fig. 1). The best fair BC protocol from among
the class of Alice-supplied BC\ protocols where $\rho _{0}$ and $\rho _{1}$
are both qubit states or at least one of $\rho _{0}$ and $\rho _{1}$ is pure
has $C^{\max }=G^{\max }=\frac{\sqrt{5}-1}{4}\simeq .\,30902$ (point B on
Fig. 1). Finally, the best fair BC protocol from among the class of
Alice-supplied BC\ protocols where $\rho _{0}$ and $\rho _{1}$ are both pure
states has $C^{\max }=G^{\max }=\frac{1}{2\sqrt{2}}\simeq .\,35355$ (point C
on Fig. 1).

\section{Significance for coin tossing}

We briefly discuss the relevance of these results to coin tossing\cite
{LCcointossing,Aharonov,KentRelCT}. Coin tossing(CT) is a cryptographic task
wherein at the end of the protocol both parties decide, based on the outcome
of their measurements, whether they have won, lost, or detected the other
party cheating. If neither party is caught cheating, then the protocol must
be such that the two parties agree on who won the coin toss. We can define a
party's {\em bias} in a CT protocol as the difference between their
probability of winning and 1/2. A CT protocol with maximum bias $\alpha $
for Alice and maximum bias $\beta $ for Bob is one where if Bob is honest,
the maximum Alice can make her probability of winning is $\frac{1}{2}+\alpha
$, and if Alice is honest, the maximum Bob can make his probability of
winning is $\frac{1}{2}+\beta .$ CT can be built upon BC as follows. After
the commitment phase, Bob sends Alice a bit which represents his guess of
her commitment. If his guess corresponds to the bit Alice unveils, he wins
the coin toss; if not, Alice wins. Our results show that it is possible to
build a secure CT protocol for any pair of biases satisfying $\alpha +\beta
\ge 1/2,$ and that this inequality can be saturated. In particular, a fair
CT protocol with both biases equal to $0.25$ can be built up in this way.

Since CT is a weaker primitive than BC\cite{KentRelCT}, the impossibility of
a BC protocol that is arbitrarily concealing and binding does {\em not}
imply the impossibility of a CT protocol with arbitrarily small biases for
both parties\cite{Localitycaveat2}. Whether such a protocol is possible
remains an open question in quantum cryptography.

It should be noted that even if such a CT protocol does not exist, the fact
that there exist CT protocols with bounded biases for both parties is still
potentially very useful. For instance, these can provide protocols for
gambling\cite{GoldenbergVaidman} wherein both parties (the casino and the
gambler) can be assured that their probability of winning is greater than
some bound, regardless of the actions of the other party.

\section{Related optimization problems}

  The central result of this paper has been the {\it maximization}
of Alice's control for certain BC protocols. However, Alice may wish to
sacrifice some control in order to reduce her probability of being caught
cheating. Specifically, if Alice assigns costs to the various outcomes of a
BC protocol, then in order to optimize her costs she must know the {\em %
minimum} probability of being caught cheating for every possible degree of
control. Since this probability quantifies the degree to which she has
`disturbed' the outcome of the protocol from what it would have been had she
been honest, we may call the result of this optimization problem the {\it %
control vs. disturbance} relation.

It is also interesting to consider a simple generalization of BC (which one
might call `integer commitment'), wherein Alice seeks to unveil one of a set
of more than two integers (rather than just `0' or `1'), and to consider the
generalization of the optimization problems mentionned above, namely, the
problems of determining the maximum probability that Alice can successfully
unveil the integer of her choosing. and the minimal probability of being
caught cheating for every possible probability of success.

These optimization problems have obvious analogies in the context of quantum
state estimation. When discriminating a set of states, one often seeks to
determine both the maximum probability of correctly estimating the state
(the maximum information gain), as well as the minimum disturbance upon the
system that is incurred for every possible degree of information gain (the
information gain vs. disturbance relation). This suggests that it may be
fruitful to pursue the analogy between the notions of control and
information gain in more detail. In future work we hope to consider these
optimization problems in the context of purification BC protocols.

\section{Conclusion}

  We have studied the extent to which BC protocols can be made
simultaneously both partially concealing and partially binding. The degrees
of concealment and bindingness were quantified by Bob's maximum information
gain about the bit committed and Alice's maximum control over the bit she
unveils. A lower bound on Alice's maximum control and Bob's maximum
information gain for {\em any }BC protocol has been derived, although it is
not known whether or not this bound can be saturated. A stronger lower bound
was obtained for a restricted class of BC protocols, called `Alice-supplied'
protocols, wherein Alice provides Bob with all of the systems that he makes
use of during the commitment phase. Moreover, this lower bound has been
shown to be saturated by what we have called a `purification' BC protocol,
wherein an honest Alice must prove her commitment to Bob by providing him
with a purification of the state she submitted to him during the commitment
phase.

We have also considered the trade-off between concealment and bindingness
for Alice-supplied BC\ protocols given different constraints on $\rho _{0}$
and $\rho _{1}$ (these are the states of the systems in Bob's possession
during the holding phase given commitments of $0$ and $1$ respectively).
Such constraints might arise from practical restrictions on the physical
implementation of a BC protocol. We have shown that for BC protocols where $%
\rho _{0}$ and $\rho _{1}$ have supports in a single 2D Hilbert space, or
wherein $\rho _{0}$ and $\rho _{1}$ are not both mixed, one cannot achieve
the optimal trade-off relation (that is, the optimal degree of bindingness
for every degree of concealment). Using protocols wherein $\rho _{0}$ and $%
\rho _{1}$ are both pure, one does even worse. The optimal trade-off for
Alice-supplied BC protocols is $C^{\max }+G^{\max }=\frac{1}{2}$ and can be
achieved using a purification BC protocol wherein $\rho _{0}$ and $\rho _{1}$
are mixed but commuting states of a 3-dimensional Hilbert space.

The following question concerning the degrees of concealment and bindingness
in BC\ protocols remains unanswered: do there exist any BC protocols with a
trade-off relation that is better than the linear trade-off relation $%
C^{\max }+G^{\max }=\frac{1}{2}?$ In order to settle this question, the
scope of our analysis must be extended beyond Alice-supplied protocols. We
conjecture that the linear trade-off is in fact the optimal trade-off from
among {\em all} BC protocols.

We end with some comments on the broader significance of the
results of this paper.   Alice's cheating strategy in a BC
protocol is an example of a task that can be described as the
preparation of quantum states at a remote location. There are many
tasks of this sort, which differ in the constraints imposed upon
the `preparer'. These constraints may specify what is known about
the state to be prepared, whether the parties involved in its
implementation are cooperative or adversarial, and how much
resource material is available, such as the number of classical or
quantum bits that can be exchanged, and the amount of prior
entanglement the parties share. For instance, in purification BC
protocols, Alice seeks to maximize her probability of remotely
preparing one of two states of a bipartite system (which may be
entangled), given that Bob is adversarial and given that she only
learns which state she wishes to prepare after she has already
submitted half of the system. (Equivalently, one may say that the
states which Alice must remotely prepare are improper mixed
states, and that she proves that she has done so by providing
purifications of these states.) There has also been interest
recently in a different sort of task involving the preparation of
quantum states at a remote location \cite{RSP}. In this task, the
parties are cooperative and the optimization problem to be solved
is the minimization of the number of classical bits of
communication asymptotically required to remotely prepare a state
for a given amount of prior entanglement. Although this task has
been called `remote state preparation', this term may be better
suited as a label for all tasks involving the preparation of
quantum states at a remote location, just as the term `state
estimation' refers to many tasks differing in the constraints
imposed on the `estimator'.

We feel that the general problem of remote state preparation may be, in some
sense, as fundamental in quantum mechanics as the general problem of state
estimation. In particular, a greater understanding of remote state
preparation may have significance for foundational research. It has been
proposed \cite{FuchsZeilinger} that the structure of quantum mechanics might
be deduced from some simple information-theoretic principles, for instance,
assumptions about how well information can be gathered, manipulated and
stored in our universe. Critical to the program is determining the extent to
which various information processing tasks can be successfully implemented
using quantum resources. The implications of our results for various
cryptographic tasks constitute a contribution to this endeavour. Ultimately
however, the program requires understanding the success of all achievable
tasks in terms of a few simple facts about information processing, for
instance, facts about a few `primitive' tasks. It has been speculated by
Fuchs that the task of state estimation is such a primitive. We add to this
our own speculation, namely, that the task of preparing quantum states at a
remote location is another such primitive.

{\it Note added.} After the completion of this research the authors were
informed\cite{Brassard} of related results obtained by A. Ambainis on fair
coin tossing protocols with bounded biases.

\section{Acknowledgments}

This work was supported by the National Science and Engineering Research
Council of Canada, the Austrian Science Foundation FWF, and the TMR programs
of the European Union Project No. ERBFMRXCT960087.

\section*{\protect  Appendix}

{\bf Proof of Lemma 1. }The density operators for qubits can be represented
by vectors on the Bloch sphere. If $\rho $ and $\sigma $ are represented by
vectors $\vec{r}$ and $\vec{s},$ then in terms of these, the trace distance
and fidelity squared can be written as \cite{NielsenChuang,Jozsa}
\begin{eqnarray*}
D(\rho ,\sigma ) &=&\frac{1}{2}\left| \vec{r}-\vec{s}\right| , \\
F(\rho ,\sigma )^2 &=&\frac{1}{2}\left( 1+\vec{r}\cdot \vec{s}+\sqrt{\left(
1-\left| \vec{r}\right| ^{2}\right) \left( 1-\left| \vec{s}\right|
^{2}\right) }\right) .
\end{eqnarray*}
Defining $r=\left| \vec{r}\right| ,s=\left| \vec{s}\right| $ and $\cos \phi =%
\vec{r}\cdot \vec{s}/rs,$ we have
\begin{eqnarray*}
D+F^{2} &=&\frac{1}{2}\sqrt{r^{2}+s^{2}-2rs\cos \phi } \\
&&+\frac{1}{2}\left( 1+rs\cos \phi +\sqrt{\left( 1-r^{2}\right) \left(
1-s^{2}\right) }\right) .
\end{eqnarray*}
This is minimized for $\phi =0.$ Moreover, assuming (arbitrarily) that $r\ge
s,$ we have $\sqrt{r^{2}+s^{2}-2rs}=r-s$ and $\sqrt{\left( 1-r^{2}\right)
\left( 1-s^{2}\right) }\ge \left( 1-r\right) \left( 1+s\right) .$ Together,
these facts imply $D(\rho ,\sigma )+F(\rho ,\sigma )^{2}\ge 1.$ $\Box $


\begin{references}
\bibitem{Localitycaveat}  A discussion of the progression toward the full
impossibility theorem can be found in \cite{Bub}. Note that we here consider
only BC\ protocols that can be implemented between parties who both have
only a single laboratory that is localized in space. Kent has shown\cite
{Kent} that unconditionally secure BC {\em can }be achieved if one relaxes
this constraint, for instance, if both parties have a pair of laboratories
separated by a distance $L$ (so that there are four laboratories in all),
each of which can be verified to be a distance $d$ to one of the other
party's laboratories, with $d\ll L.$ The security stems from the constraint
imposed by special relativity on the speed of communications between a
party's two laboratories. Unfortunately, the protocol requires a channel
capacity that increases exponentially with the duration of the holding
phase. See \cite{Bub} for further discussion.
\bibitem{Localitycaveat2}  Again, we are only considering protocols that can
be implemented between localized parties, as discussed in Ref.\cite
{Localitycaveat}. It has been shown by Kent\cite{KentRelCT} that if one
relaxes this constraint, unconditionally secure coin tossing with
arbitrarily small biases can be achieved.
\end{references}
\end{document}